\pdfoutput=1

\documentclass[11pt]{article}

\usepackage{ACL2023}

\usepackage{times}
\usepackage{latexsym}

\usepackage[T1]{fontenc}

\usepackage[utf8]{inputenc}

\usepackage{microtype}

\usepackage{inconsolata}

\usepackage{graphicx}
\usepackage{booktabs}
\usepackage[title]{appendix}

\title{On Barriers to Archival Audio Processing}

\author{Peter Sullivan \\
  University of British Columbia, Canada \\
  \texttt{prsull@student.ubc.ca} \\\And
  Muhammad Abdul-Mageed \\
  University of British Columbia, Canada \\
  \texttt{muhammad.mageed@ubc.ca} \\}

\begin{document}
\maketitle
\begin{abstract}
In this study, we leverage a unique UNESCO collection of mid-20th century radio recordings to probe the robustness of modern off-the-shelf language identification (LID) and speaker recognition (SR) methods, especially with respect to the impact of multilingual speakers and cross-age recordings. Our findings suggest that LID systems, such as Whisper, are increasingly adept at handling second-language and accented speech. However, speaker embeddings remain a fragile component of speech processing pipelines that is prone to biases related to the channel, age, and language. Issues which will need to be overcome should archives aim to employ SR methods for speaker indexing.   
\end{abstract}

\section{Introduction}
Multinational organizations such as the United Nations (UN); the International Federation of Red Cross and Red Crescent Societies (IFRC); and the United Nations Educational, Scientific, and Cultural Organization (UNESCO) maintain audio archives that are of profound cultural and historical value. However, incomplete descriptive metadata often hinder their access by the public~\cite{zervanou2011enrichment}.

Audio archives present a complex terrain for contemporary speech processing technologies, owing to the varied domains these recordings encapsulate. The long running MALACH project's effort to tackle emotional, disfluent, and accented speech~\cite{picheny19_interspeech,psutka2002automatic} gives of sense of this complexity. In this study, our attention is directed towards extensive multilingual repositories, which pose challenges for speaker recognition (SR) and language identification (LID) technologies due their long chronological span and inclusion of second-language (L2) speech.

A primary objective in enhancing accessibility to these recordings involves identifying the speakers within a specific recording. While documentation of speakers is \textit{sometimes} available, it is only at the document level, making this task closer to speaker indexing~\cite{sturim2001speaker} than standard SR. 
Moreover, concerns regarding the robustness of speaker embeddings to voice modifications associated with aging, as well as the accuracy of language-specific predictions~\cite{hutiri2022bias}, significantly challenge the straightforward utilization of off-the-shelf SR technologies within the ambit of these long-running, multilingual archives.
    
     
In this exploratory investigation using a selection of radio audio archives from UNESCO, we explore the impact of these factors on robustness of zero-shot application of off-the-shelf tools, to identify paths towards speaker indexing in age- and language-variable environments. 
Our dataset involves 171 hours of archival data covering the period of 1952-1980, involving 20 languages (See Table \ref{tab:data}). Our work offers the following contributions:
    \begin{enumerate}
        \item We characterize a relatively neglected sphere within speech processing scholarship: multilingual audio archives.
        \item We carry out a cross-age analysis to investigate robustness of the speaker embeddings.
        \item We analyze the robustness of speaker embeddings in multilingual speech scenarios, uncovering unique insights that are otherwise hidden without access to datasets such as ours.
        \item We investigate the utility of a number of off-the-shelf language identification tools for accented LID.
    \end{enumerate}
    
    The rest of the paper is organized as follows: Section~\ref{section:background} offers a brief overview of speech processing as it relates to challenges present in archival audio. Section~\ref{section:datasets} is a description of our datasets. Section~\ref{section:dataprocessing} details our data processing methods. Section~\ref{section:models} gives an overview of the LID and SR models used. Section~\ref{section:experiments} is a breakdown of our different experiments. In Section~\ref{section:results}, we provide our experimental results and discuss these in Section~\ref{section:discussion}. We discuss limitations of our work in Section~\ref{section:limitations} and provide ethical implications in Section~\ref{section:ethics}. We conclude in Section~\ref{section:conclusion}.

\section{Background}
\label{section:background}

\subsection{Language ID}
Recent work in LID has moved from the discrete segment representations popular in earlier i-vector~\cite{dehak11_interspeech} and x-vector~\cite{snyder18_odyssey} works, to the convenience of end-to-end deep neural models either based on ResNets~\cite{cai2018novel} or Transformers~\cite{babu2021xls,radford2023robust,pratap2023scaling}. Part of this has been enabled by the creation of LID datasets such as VoxLingua107~\cite{valk2021voxlingua107} and FLEURS~\cite{conneau2023fleurs}, allowing for direct training of large end-to-end models. Meanwhile, another aspect of this evolution has been the inclusion of LID into models designed as `jack-of-all-trades' tools such as Whisper~\cite{radford2023robust} and Massively Multilingual Speech (MMS)~\cite{pratap2023scaling}, which are built to support automatic speech recognition (ASR) and speech translation in addition to LID. 

Recent work in LID has raised attention to less well developed areas of exploration including L2 LID~\cite{kukk22_interspeech}, LID for multilingual users~\cite{titus2020improving}, and domain generalization of LID~\cite{sullivan23_interspeech}. For archival audio, these three aspects become even more relevant, as the long chronological nature of the archives and diverse population of multilingual speakers demand the use of highly robust LID systems. At the same time, the operational context (including lack of funding and know-how) means that off-the-shelf tools are the only viable option, neglecting transfer learning or domain adaptation approaches, regardless of their efficacy.

\subsection{Speaker embedding}

Modern SR makes use of speaker embeddings (i.e. x-vector, r-vector etc.) often extracted from ResNets-based extractor networks ~\cite{desplanques20_interspeech,wang2023wespeaker} and trained on VoxCeleb~\cite{nagrani17_interspeech}. 
Despite the progress made in the field, a recent study of bias in SR~\cite{hutiri2022bias} indicate a number of issues, including demographic biases impacting age, gender, language, and nationality. To our knowledge little work has been performed to investigate the impact of language (including multilinguality) on SR effectiveness. However, there is a body of work that has looked at the impact of cross-age scenarios on SR~\cite{singh23d_interspeech,qin22_interspeech}. With \citet{qin22_interspeech} employing adversarial learning to learn age-invariant embeddings, and \citet{singh23d_interspeech} estimating the impact of age on cross-age verification. To our knowledge no work has considered investigation of cross-age scenarios in multilingual corpora.

Work in speaker indexing, has been largely neglected since anchor model based approaches~\cite{sturim2001speaker}, but we hope that this study will pave the way to future work in speaker indexing by identifying existing obstacles with off-the-shelf tools.


\section{Datasets}
\label{section:datasets}

\textbf{VoxPopuli - L2 English}
An existing source of L2 audio, in a similar domain to the one under consideration is the VoxPopuli collection of EuroParl recordings~\cite{wang2021voxpopuli}. In particular we are interested in the 29 hour subset of L2 English speech coming from 15 different accent backgrounds.

\textbf{Radio Archives}
From our partnership with UNESCO, we utilize a mid-20th century radio recording archives collection that is currently undergoing metadata enrichment. While the dataset is currently not published, the metadata used and links to the recordings will be made available. The recordings in this collection consist of a variety of material including speeches, interviews, news briefs, and educational programs. The audio was recorded between the 1950s and early 1980s, and digitized in the late 2010s. See Table \ref{tab:data} for more detailed statistics.

\begin{table}
  \caption{Dataset Overview. n is the number of recordings. For LID, we use only the first 30 seconds of spoken audio from each recording.}
  \label{tab:data}
  \centering
  \begin{tabular}{lcccc}
    \toprule
   Split & n & Hours & Languages\\
    \midrule
    LID & $484$ & $4$ & $20$ \\
    Cross-age & $692$ & $104$ & $17$ \\
    Cross-lingual & $463$ & $67.5$ & $20$ \\
      
    \bottomrule
  \end{tabular}
\end{table}

\section{Data Processing}
\label{section:dataprocessing}
While the VoxPopuli dataset is used with existing segmentation for our L2 English experiments, we build two datasets from the radio archives by filtering the known metadata to meet certain criteria. As much of the radio archives contain incomplete metadata, we restrict our selection to recordings identified as having a single known speaker on the recording,
as well as a single languages spoken. 

From this selection, we create two datasets: Our \textit{cross-age} dataset is filtered by selecting speakers with multiple recordings in the same language occurring in different calendar years.
The second \textit{cross-lingual} dataset is filtered by selecting speakers with multiple recordings in different languages. 

For both datasets, we further filter by diarizing the recordings using Pyannote's~\cite{bredin2020pyannote} speaker diarization pipeline (version 3.1), and selecting recordings where there is a single speaker accounting for more than $75\%$ of the duration. We utilize 16khz single channel copies of the recordings for the study.

\section{Models}
\label{section:models}
\textbf{Whisper}~\cite{radford2023robust} is a multilingual speech processing model that allows for ASR, speech translation, and LID. We use all three versions of the large (1.5B parameter) model.

\noindent\textbf{MMS}~\cite{pratap2023scaling} is similarly a multilingual speech processing model allowing for ASR, speech translation, LID, and additionally text-to-speech. The model has been primarily trained on the Bible and other religious audio, with a focus on scaling the number of covered languages. We use the (1B parameter) 126 language version.

\noindent\textbf{WeSpeaker ResNet34-LM}~\cite{wang2023wespeaker} is a SR model trained on VoxCeleb~\cite{nagrani17_interspeech}.

\section{Experiments}
\label{section:experiments} 
\subsection{L2 LID}
For our L2 LID experiments, we compare two well known off-the-shelf LID models: Whisper Large V(1-3)~\cite{radford2023robust} and MMS L126~\cite{pratap2023scaling}. For our VoxPopuli dataset we simply report predictions on the given segments. However, for our cross-lingual radio dataset, we follow the suggested LID procedure of the Whisper model and select the first 30 seconds of audio to perform the prediction.

\subsection{Speaker Embedding Robustness}
To understand how robust speaker embeddings are to cross-age and cross-lingual effects, we use a pretrained ResNet model, Wespeaker ResNet34-LM~\cite{wang2023wespeaker}, to extract segment representations. We take the average embedding for the majority speaker in each recording. For the \textit{cross-age experiments}, we calculate the cosine similarity between representations and aggregate by difference in calendar years between recordings. We rely on calendar year as an approximation as we do not have recording dates for some recordings. For the \textit{cross-lingual experiments}, we simply compare the cosine similarity between same language segments to their different language similarity scores. 

\section{Results}
\label{section:results}
For L2 English performance (see Table \ref{tab:voxpoluli_l2} in Appendix \ref{appendix:lid}), 
The updated Whisper V3 model substantially improves the recognition of accented English audio attaining an accuracy of $94\%$ compared to both Whisper V1,V2 and MMS L126.

For our mixed L1 and L2 multilingual archival audio (see Table \ref{tab:radio_lid} in Appendix \ref{appendix:lid}), we see similar results as to the VoxPopuli results. Notably, MMS performs better on the mixture of accented and unaccented speech (accuracy at 71.90\%). Still, the Whisper V3 model performs best on this set of audio: accuracies of 88.01\% for V1, 87.60 for V2, and 91.32\% for V3.

\begin{figure}[t]
  \centering
  \includegraphics[width=\linewidth]{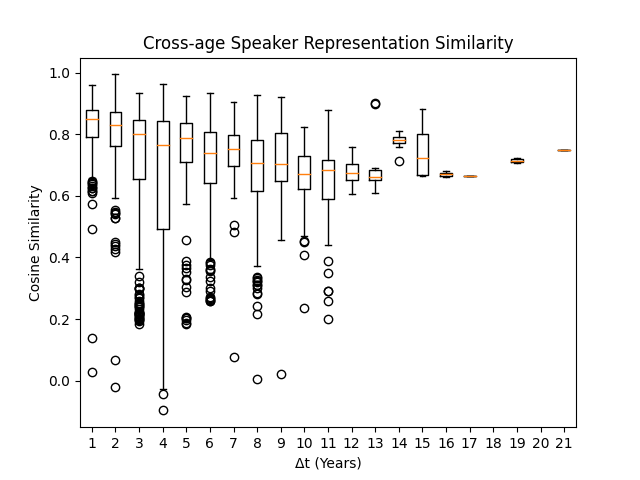}
  \caption{Longitudinal comparison of speaker embedding. Cosine similarity scores are plotted with median and quartiles marked. Outliers are noted as circles. $\Delta t$ is the absolute difference in calendar years between each pair of recordings. Data past 15 years becomes quite sparse, as few speakers fit our filtering criteria.} 
  \label{fig:longitudinal_comparison}
\end{figure}

\begin{figure}[t]
  \centering
  \includegraphics[width=\linewidth]{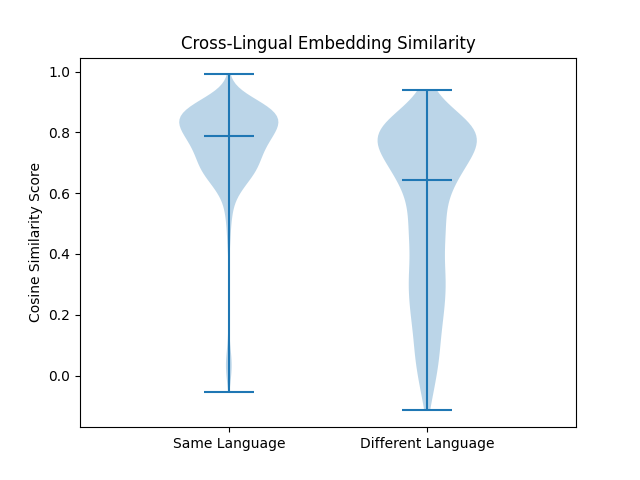}
  \caption{Cross-lingual comparison of speaker embedding cosine similarity scores. To better show the distribution, we present the results as violin plots, noting that the lower end of the Same Language plot may be representative of different speakers who were not filtered out of our automated approach. Distribution statistics: \textit{Same language} Mean: $0.71$, Median: $0.76$, Std: $0.19$; \textit{Different language} Mean: $0.53$, Median: $0.60$, Std: $0.26$}
  \label{fig:cross_lingual_comparison}
\end{figure}

Looking at the robustness of the speaker embeddings, we notice a substantial drop in the similarity scores in both the cross-age setting (see Fig. \ref{fig:longitudinal_comparison}) and cross-lingual setting (see Fig. \ref{fig:cross_lingual_comparison}). For the cross-age comparison, median cosine similarity scores continue to drop until stabilizing after a gap of 10 years. While cross-lingual performance of representations drops substantially as well, of note is the very large increase in standard deviation between the two settings. This latter trend is potentially problematic for treating all cross-lingual scenarios the same, and may be indicative that fluency as well as language similarity between compared languages may be factors.

\section{Discussion}
\label{section:discussion}
The performance of the Whisper V3 model appears to indicate it as a strong candidate for archives in processing multilingual audio files. Notably, while all of the Whisper models under examination have roughly the same number of parameters (V3 is slightly larger due to increasing the input dimensions), the larger amount of training data used for V3 appear to substantially help it in identifying accented English. The brittle nature of the MMS model with regard to L2 speech is quite surprising, and demonstrates the importance of having diverse and challenging audio benchmarks for LID. While the longitudinal embedding comparison demonstrates the challenges of applying SR models across channel and age. The appearance of a flattening at around 10 years of difference indicates that it may be possible to account for this cross-age drift.

\section{Limitations}
\label{section:limitations}
Working with real world datasets presents substantial challenges and limitations. While the archival partner aims to put the radio recordings online sometime in the future, these are not yet available publicly. Similarly, working with aggregated data has limitations. For instance the small cluster of low similarity in the same language embedding comparison is likely indicative of misidentified speakers who slipped through the filtering process. 

\section{Ethics}
\label{section:ethics}
Speaker identification tools have the potential for significant privacy violation, especially if applied as part of a surveillance system at scale. As seen from the study, the large cross-channel and cross language variability in speaker embeddings present significant risks for use in larger scales, where risk associated with misidentification may also be quite high.

We believe our investigation of these tools for the purpose of improving archival accessibility and discovery are consistent with ethical practice, and our application to a limited closed set of public figures on public audio recordings further limits this potential for abuse. 

\section{Conclusion}
\label{section:conclusion}
We present an analysis into the robustness of SR and LID tools on cross-age, and cross-lingual audio on a unique archival radio dataset. Our work indicates the viability of existing LID tools, such as Whisper V3~\cite{radford2023robust}, at handling the accented and multilingual speech common in recordings from multinational organizations. 
We also find that cross-age and cross-lingual application of SR introduce large drops in the cosine similarity scores, with the cross-age similarity dropping steadily over a period of a decade. The cross-lingual scores further introducing large increase in the standard deviation, potentially indicative of other factors such as language fluency or language similarity impacting the result. Additionally, this work demonstrates the value of archival audio in examining current speech processing approaches. The variety of such recordings offer a strong platform to study bias and domain adaptation.


\section{Acknowledgments}
This work was performed as part of the Interpares Trust AI research project, focusing on AI by and for Archives.

\bibliography{anthology,custom}
\bibliographystyle{acl_natbib}

\clearpage
\begin{appendices}

\section{LID Experiments}
\label{appendix:lid}

\begin{table}[h]
  \caption{L2 English LID Performance of Whisper~\cite{radford2023robust} and MMS~\cite{pratap2023scaling}}
  \label{tab:voxpoluli_l2}
  \centering
  \begin{tabular}{ll}
    \toprule
    \textbf{Model}      & \textbf{Accuracy}                \\
    \midrule
    Whisper-Large v1 
    & $72.65\%$     \\  
    Whisper-Large v2
    & $72.65\%$     \\  
    Whisper-Large v3
    & $94.52\%$  \\   MMS L126
    & $11.10\%$
    \\  
    \bottomrule
  \end{tabular}
\end{table}

\begin{table}[h]
  \caption{Multilingual Archival Audio LID Performance of Whisper~\cite{radford2023robust} and MMS~\cite{pratap2023scaling}}
  \label{tab:radio_lid}
  \centering
  \begin{tabular}{ll}
    \toprule
    \textbf{Model}      & \textbf{Accuracy}                \\
    \midrule
    Whisper-Large v1 
    & $88.01\%$     \\  
    Whisper-Large v2 
    & $87.60\%$     \\  
    Whisper-Large v3 
    & $91.32\%$  \\ 
    MMS L126 
    & $71.90\%$
    \\  
    \bottomrule
  \end{tabular}
\end{table}
\end{appendices}

\end{document}